\documentclass[a4paper]{article}
\usepackage[utf8]{inputenc}

\usepackage{amsmath}
\usepackage{amssymb}
\usepackage{amstext}
\usepackage{mathtools}
\usepackage{graphicx}
\usepackage{fullpage}
\usepackage{array}
\usepackage{tabularx}
\usepackage{multirow}
\usepackage{longtable}
\usepackage{xcolor}
\usepackage{float}
\usepackage{listings}
\usepackage{booktabs}
\usepackage{algorithm}%
\usepackage{algorithmicx}%
\usepackage{algpseudocode}%
\usepackage{array}
\usepackage{graphicx} %
\usepackage{subcaption}
\usepackage{authblk}

\makeatletter
   
\newcommand{\ALOOP}[1]{\ALC@it\algorithmicloop\ #1%
  \begin{ALC@loop}}
\newcommand{\ENDALOOP}{\end{ALC@loop}\ALC@it\algorithmicendloop}

\newcommand{\algorithmicbreak}{\textbf{break}}

\makeatother
\usepackage{listings}%

\newcommand{\bs}[1]{\boldsymbol{#1}}
\def\bsb{\bs{b}}

\def\bsv{\bs{v}}
\providecommand{\keywords}[1]
{
  \small	
  \textbf{\textit{Keywords--}} #1
}

\begin{document}

\title{Finite Element Method for HJB in Option Pricing with Stock Borrowing Fees}

\author[1]{Rakhymzhan Kazbek} 
\author[1]{Aidana Abdukarimova}

\affil[1]{Astana IT University, Department of Computation and Data Science, Mangilik El C1, Astana, Kazakhstan}

\maketitle

\begin{abstract}
In mathematical finance, many derivatives from markets with frictions can be formulated as optimal control problems in the HJB framework. Analytical optimal control can result in highly nonlinear PDEs, which might yield unstable numerical results. Accurate and convergent numerical schemes are essential to leverage the benefits of the hedging process. In this study, we apply a finite element approach with a non-uniform mesh for the task of option pricing with stock borrowing fees, leading to an HJB equation that bypasses analytical optimal control in favor of direct PDE discretization. The time integration employs the theta-scheme, with initial modifications following Rannacher`s procedure. A Newton-type algorithm is applied to address the penalty-like term at each time step. Numerical experiments are conducted, demonstrating consistency with a benchmark problem and showing a strong match. The CPU time needed to reach the desired results favors P2-FEM over FDM and linear P1-FEM, with P2-FEM displaying superior convergence. This paper presents an efficient alternative framework for the HJB problem and contributes to the literature by introducing a finite element method (FEM)-based solution for HJB applications in mathematical finance.

\end{abstract}

\section{Introduction}
Many queries from the financial derivatives market can be formulated as optimal control problems, leading to HJB-type nonlinear PDEs. However, the common approach in the literature is to obtain the analytical control and substitute it into the PDE, making the resulting HJB highly nonlinear and, hence, less amenable to solving it by discrete convergence guaranteed numerical techniques. Incorporation of the stock borrowing fees is keen to realize the short selling which includes those borrowed fees. Hence, the model reflects the market friction more reliably. 

Optimal control problems, reformulated as HJB PDE are fully nonlinear PDEs consisting the nonlinearity in first-order derivative. Usually, the treatment of degenerate or convection dominated Black-Scholes PDEs are tedious to implement and generalize for different tasks seen in Black-Scoles PDEs~\cite{Forsyth1999}. There are several well-known methods (e.g., the Discontinuous Galerkin method~\cite{kozpinar2020pricing}, Van Leer flux limiters~\cite{forsyth2002quadratic, Forsyth1999}, etc.) to control the flux of the convection term which helps to mitigate the possible instabilities. As the total variation diminishing (TVD) condition is important to prevent the amplifying possible spurious oscillations. However, the high Peclet number may introduce spurious oscillations and non monotone convergence to the desired solution, even for linear problems. One possible way of mitigating the is to avoid the variable coefficients using the Landau transformation~\cite{kazbek2024isogeometric}. This helps to control the area of possible instability as time goes on on pay-off function. The kink or any type of non-differentiability seen in pay-off functions may lead to non-monotonic convergence or even to oscillations~\cite{christara2018analysis}. The monotonic convergence can be secured by using advanced iteration techniques, such as policy or penalty-like methods~\cite{christara2022penalty,Forsyth2007}. Moreover, the Rannacher approach~\cite{Rannacher1984,christara2018analysis} is another method which helps to smooth and reduce the possible errors coming from irregular initial data. This can be associated with the nature of any dynamic problem, the errors will contribute to the numerical result from both temporal and spatial discretization. 

In part, the policy iteration will spend notably less computational time than the classical Newton-Raphson method~\cite{kazbek2024pricing} for pricing American-type contracts. The standard stopping criterias from Newton's method leads to high number of iterations to converge per each discretized time grid. For HJB PDEs the policy iteration combined with the penalty iteration is the choice to be solved, see e.g.~\cite{forsyth2002quadratic,Forsyth2007}. The policy iteration for European type HJB PDEs was revisited and improved in~\cite{christara2022penalty} with a newer stopping criterias which led to monotonic and fast convergence using relatively less amount of iterations for highly nonlinear tasks. 

The literature on HJB problems from numerous applications using FEM~\cite{HJB-FEM1,HJB-FEM2,HJB-FEM3,HJB-FEM4} is vast. However, there are no viable FEM approaches in literature dedicated to nonlinear derivative pricing models from the market with friction. To fill this gap, we propose a FEM approach to deal with HJB task for stock borrowing/lending fees (SBF) for Straddle European option~\cite{Forsyth2007}.

In previous paragraphs, we provide notable obstacles encountered in HJB PDE and, in general, Black-Scholes PDEs. To quantify and resolve abovementioned issues, we suggest a hybrid framework of variable mesh creation by FEM and Crank-Nicolson-Rannacher (CNR) techniques. We, in particular, resort to higher order FEM discretization which results in penta-diagonal matrices. The discretization of the HJB PDE is done directly without the involvement of analytical optimal control. Previously, the penalty-based (no derivatives involved in the penalty term) approach for the Convertible Bonds model~\cite{kazbek2024pricing} is solved using the FEM approach demonstrating the notable performance over classical FDM. Inspired by this we revisit the HJB PDE  with a penalty-like term (without penalty parameter). Unlike that method, we employ the non-uniform mesh without a posteriori error indicator which does not rebuilt the mesh but set it up once at initial stage to capture the possible error outcomes from the discretization. Moreover, the post-processing Greek values are of crucial importance at least as the price of the contracts. We also provide a comparison between FEM and FDM-based Greek results.

The contribution of the paper to the computational derivative pricing field is a triplet. A promising FEM alternative approach is presented with a visible advantages over FDM. Relative execution timing strengthens the motivation to use for practical purposes. Lastly, the Greeks are provided in a consistent way inheriting the advantages shown in the valuation and then transferring to Greek values. 

Section 2 introduces the HJB models for a European straddle option incorporating stock borrowing fees. Section 3 outlines the spatial and temporal discretization methods that result in a solvable system of equations. The efficiency of the proposed framework is then demonstrated through numerical experiments in Section 4. Finally, concluding remarks are provided in Section 5.

\section{Model formulation}\label{sec: HJB model}
We are interested in HJB PDEs for short and long-position problems for SBF. The price of the European straddle option with SBF~\cite{Forsyth2007}, at any time $t\in(0,T)$ and the underlying variable $S\in(S_{min},S_{max})$, can be computed by solving the discrete optimal control problem:
\begin{equation}
    V_\tau=\sup _Q\left\{\frac{\sigma^2 S^2}{2} V_{S S}+q_3 q_1\left(S V_S-V\right)+\left(1-q_3\right)\left(\left(r_l-r_f\right) S V_S-q_2 V\right)\right\}
\end{equation}
with $Q=\left(q_1, q_2, q_3\right), q_1 \in\left\{r_l, r_b\right\}, q_2 \in\left\{r_l, r_b\right\}, q_3 \in\{0,1\}$ for the short position. For the long position, the sup shall be replaced by inf. 
The corresponding nonlinear HJB PDE with terminal and boundary conditions are as follows
\begin{equation}\label{eq:HJB original short}
    \begin{cases}\displaystyle 
     \frac{\partial V}{\partial t}=\frac{\sigma^2 S^2}{2} \frac{\partial^2 V}{\partial S^2}+r_l\left(S \frac{\partial V}{\partial S}-V\right)+\max \left\{\left(r_b-r_l\right)\left(S \frac{\partial V}{\partial S}-V\right),-r_f S \frac{\partial V}{\partial S}, 0\right\}~~~\text{in $\Omega$}.  \\ 
    V(S,T)=\max(S-K,K-S), \\ 
    \lim _{S \rightarrow \infty} V(S, t)\approx S-K~~~\text{on $\Gamma_D$}, \\ 
    \lim _{S \rightarrow -\infty} V(x,t)=K-S~~~\text{on $\Gamma_D$},
    \end{cases}
\end{equation}
for the short position and
\begin{equation}\label{eq:HJB original long}
    \begin{cases}\displaystyle 
     \frac{\partial V}{\partial t} + \frac{\sigma^2 S^2}{2} \frac{\partial^2 V}{\partial S^2}+r_bS \frac{\partial V}{\partial S} - r_bV+\min \left\{\left(r_l-r_b\right)\left(S \frac{\partial V}{\partial S}-V\right),-\left(r_b-r_l+r_f\right) S \frac{\partial V}{\partial S}, 0\right\}~~~\text{in $\Omega$}.  \\ 
    V(S,T)=\max(S-K,K-S), \\ 
    \lim _{S \rightarrow \infty} V(S, t)\approx S-K~~~\text{on $\Gamma_D$}, \\ 
    \lim _{S \rightarrow -\infty} V(x,t)=K-S~~~\text{on $\Gamma_D$},
    \end{cases}
\end{equation}
for the long position. 

The classical time reversion transformation is applied, as we solve the problem based on the terminal pay-off function. The variable coefficients are reduced into constant ones by using the Landau~\cite{Zhu_Landau} transformation. It leads to the analytical integration results without the use of quadrature techniques in FEM context. Applying the change of variables results in the transformed HJB PDEs for the short and long positions, respectively. 
\begin{itemize}
    \item $\tau = T-t$
    \item $ x = \log(S/K)$
\end{itemize} 
\begin{equation}\label{eq:HJB_transformed short}
    \begin{cases}\displaystyle 
    \frac{\partial V}{\partial \tau} = \frac{\sigma^2}{2} \frac{\partial^2 V}{\partial x^2}+\left(r_b - \frac{\sigma^2}{2}\right) \frac{\partial V}{\partial x} - r_bV+\max \left\{\left(r_b-r_l\right)\left( \frac{\partial V}{\partial x}-V\right),-r_f  \frac{\partial V}{\partial x}, 0\right\}~~~\text{in $\Omega$}.  \\ 
    V(x,0)=\max(Ke^x-K,K-Ke^x), \\ 
    \lim _{x \rightarrow \infty} V(x, \tau)\approx Ke^x-K~~~\text{on $\Gamma_D$}, \\ 
    \lim _{x \rightarrow -\infty} V(x,\tau)=K-Ke^x~~~\text{on $\Gamma_D$},
    \end{cases}
\end{equation}
and
\begin{equation}\label{eq:HJB_transformed long}
    \begin{cases}\displaystyle 
    \frac{\partial V}{\partial \tau} = \frac{\sigma^2}{2} \frac{\partial^2 V}{\partial x^2}+\left(r_b - \frac{\sigma^2}{2}\right) \frac{\partial V}{\partial x} - r_bV+\min \left\{\left(r_l-r_b\right)\left( \frac{\partial V}{\partial x}-V\right),-\left(r_b-r_l+r_f\right)  \frac{\partial V}{\partial x}, 0\right\}~~~\text{in $\Omega$}.  \\ 
    V(x,0)=\max(Ke^x-K,K-Ke^x), \\ 
    \lim _{x \rightarrow \infty} V(x, \tau)\approx Ke^x-K~~~\text{on $\Gamma_D$}, \\ 
    \lim _{x \rightarrow -\infty} V(x,\tau)=K-Ke^x~~~\text{on $\Gamma_D$},
    \end{cases}
\end{equation}
where $\Gamma_D = \Gamma \equiv \partial\Omega$. The change of variables are applied for the short position in the same manner.  
Let us denote $\displaystyle\left(r_b-r_l\right)\left( \frac{\partial V}{\partial x}-V\right) = \mathsf{S}_1 V$  and $\displaystyle-r_f  \frac{\partial V}{\partial x} = \displaystyle\mathsf{S}_2 V$ in the $\max$ term  and $\displaystyle(r_l-r_b)\left( \frac{\partial V}{\partial x}-V\right) = \mathsf{L}_1 V$  and $\displaystyle-\left(r_b-r_l+r_f\right)  \frac{\partial V}{\partial x} = \displaystyle\mathsf{L}_2 V$ in the $\min$ term . The nonlinear sources in the~\eqref{eq:HJB_transformed long} and \eqref{eq:HJB_transformed short}, can be rewritten for the convenience as:
\begin{equation}\label{eq:HJB for weak form. short}
    \displaystyle 
    \frac{\partial V}{\partial \tau} = \frac{\sigma^2}{2} \frac{\partial^2 V}{\partial x^2}+\left(r_b - \frac{\sigma^2}{2}\right) \frac{\partial V}{\partial x} - r_bV + \alpha_{short},§
\end{equation}
and 
\begin{equation}\label{eq:HJB for weak form. long}
    \displaystyle 
    \frac{\partial V}{\partial \tau} = \frac{\sigma^2}{2} \frac{\partial^2 V}{\partial x^2}+\left(r_b - \frac{\sigma^2}{2}\right) \frac{\partial V}{\partial x} - r_bV + \alpha_{long},§
\end{equation}
where 
\begin{eqnarray}\label{eq: alpha functions for short and long}
    \alpha_{short}= \begin{cases}0 & \text { if }\mathsf{S}_1 V\leq 0 \text { and }\mathsf{S}_2 V\leq 0 \\
\mathsf{S}_1 V & \text { if }\mathsf{S}_1 V >0 \text { and }\mathsf{S}_1 V>\mathsf{S}_2 V \\ 
\mathsf{S}_2 V & \text { if }\mathsf{S}_2 V>0 \text { and }\mathsf{S}_1 V \leq\mathsf{S}_2 V .\end{cases} \quad \text{and} \quad
\alpha_{long}= \begin{cases}0 & \text { if }\mathsf{L}_1 V\geq 0 \text { and }\mathsf{L}_2 V\geq 0 \\
\mathsf{L}_1 V & \text { if }\mathsf{L}_1 V <0 \text { and }\mathsf{L}_1 V<\mathsf{L}_2 V \\ 
\mathsf{L}_2 V & \text { if }\mathsf{L}_2 V<0 \text { and }\mathsf{L}_1 V \geq\mathsf{L}_2 V .\end{cases}
\end{eqnarray}

\section{Methodology}

\subsection{Finite Element Method}\label{sec: Finite element method}
As in any family of FEM-based methods, the numerical solution of the PDEs~\eqref{eq:HJB_transformed long}, are defined in the weak sense. Hence we introduce the the Sobolev space $\mathcal{H}^1(\Omega)$, where the $V$ is defined in the weak sense:
$$
\mathcal{H}^1(\Omega)=\left\{f\left| f,~\frac{\partial f}{\partial x} \in L^2(\Omega),\right| \right\} .
$$
The next step is to define the space for trial solutions $\mathcal{S}$, the requirement to be square integrable functions and satisfaction of Dirichlet boundary condition:
$$
\mathcal{S}=\left\{f\left|f \in \mathcal{H}^1(\Omega), f\right|_{\Gamma_D}=g\right\} .
$$
Then the space for the class of weighting functions $\mathcal{V}$ are defined to be vanished (homogeneous) at the boundaries as
$$
\mathcal{V}=\left\{z\left|z \in \mathcal{H}^1(\Omega), z\right|_{\Gamma_D}=0\right\} .
$$
To be concise, we only consider the HJB for the long position~\eqref{eq:HJB for weak form. long} as the second case for the short position~\eqref{eq:HJB for weak form. short} could be done in the same pattern. At the same time integration details are presented, henceforth, the numerical methodology is presented for the long position unless otherwise specified. A finite element solution of~\eqref{eq:HJB for weak form. long} can be obtained via weak formulation of~\eqref{eq:HJB for weak form. long} and is equivalent of finding the $V \in \mathcal{S}$ using the arbitrary $z \in \mathcal{V}$.
\begin{equation}
\begin{aligned}
   \int\displaylimits_{\Omega}z\frac{\partial V}{\partial \tau}dx   &= \frac{\sigma^2}{2}\int\displaylimits_{\Omega}z\frac{\partial^2 V}{\partial x^2}dx + \left(r_b- \frac{\sigma^2}{2}\right)\int\displaylimits_{\Omega}z\frac{\partial V}{\partial x}dx - r_b\int\displaylimits_{\Omega}zVdx + \int\displaylimits_{\Omega}z\alpha_{long}dx, \notag
\end{aligned}
\end{equation}
after using the product rule, the divergence theorem and applying the homogeneous property of the function $z$ at the boundary results in the weak formulation
\begin{equation}
\begin{aligned}
   \frac{\partial }{\partial \tau}\int\displaylimits_{\Omega}zV dx &= -\frac{\sigma^2}{2}\int\displaylimits_{\Omega}\frac{\partial z}{\partial x}\frac{\partial V}{\partial x}dx - \left(r_b- \frac{\sigma^2}{2}\right)\int\displaylimits_{\Omega}\frac{\partial z}{\partial x}Vdx - r_b\int\displaylimits_{\Omega}zVdx + \int\displaylimits_{\Omega}z\alpha_{long}dx, ~~~\forall z\in \mathcal{V}. \notag
\end{aligned}
\end{equation}
To build the finite-element approximation~\cite{elman2014finite}, consider the finite-dimensional subspace $S^h_0 \subset \mathcal{H}^1_0$, spanned by the basis $\{\psi_1,\psi_2,\dots,\psi_n\}$. The finite-element approximation to the solution $V$ is the function
$$
  V_h = \sum_{i=1}^n v_i \psi_i + \sum_{i \in \mathcal{I}_{\partial}} v_i \psi_i \simeq V,  \quad v_i \in \mathbb{R}.
$$
where $\psi_{i \in \mathcal{I}_{\partial}}$ are additional functions needed to interpolate the given solutions at the boundaries. The use of the above approximations results in the weak formulation in the finite-dimensional space:
\begin{equation}
\begin{aligned}
\frac{\partial}{\partial \tau} \left( \sum_{i=1}^n v_i \int\displaylimits_{\Omega} z \psi_i dx + \sum_{i\in \mathcal{I}_{\partial}} v_i \int\displaylimits_{\Omega} z \psi_i dx \right) &= -\frac{\sigma^2}{2} \left( \sum_{i=1}^n v_i  \int\displaylimits_{\Omega}\frac{\partial z}{\partial x}\frac{\partial \psi_i}{\partial x} dx + \sum_{i\in\mathcal{I}_{\partial}} v_i  \int\displaylimits_{\Omega}\frac{\partial z}{\partial x}\frac{\partial \psi_i}{\partial x} dx\right) \\
&- \left(r_b- \frac{\sigma^2}{2}\right) \left( \sum_{i=1}^n v_i \int\displaylimits_{\Omega}\frac{\partial z}{\partial x}\psi_idx   + \sum_{i\in\mathcal{I}_{\partial}}^n v_i \int\displaylimits_{\Omega}\frac{\partial z}{\partial x}\psi_i dx \right) \\
&-r_b \left( \sum_{i=1}^n v_i \int\displaylimits_{\Omega}z\psi_i dx + \sum_{i\in \mathcal{I}_{\partial}} v_i \int\displaylimits_{\Omega}z\psi_i dx \right) + \mathcal{P}_{long}.
\end{aligned}
\end{equation}
In the Galerkin method~\cite{elman2014finite}, the test function $z$ is chosen same as the basis function $\psi_i$. Applying this condition for $z = \psi_j$, $j=1,\dots,n$ gives the system of equations
\begin{equation}
\begin{aligned}
\frac{\partial}{\partial \tau} \left( \sum_{i=1}^n v_i \int\displaylimits_{\Omega} \psi_j \psi_i dx + \sum_{i\in \mathcal{I}_{\partial}} v_i \int\displaylimits_{\Omega} \psi_j \psi_i dx \right) &= -\frac{\sigma^2}{2} \left( \sum_{i=1}^n v_i  \int\displaylimits_{\Omega}\frac{\partial \psi_j}{\partial x}\frac{\partial \psi_i}{\partial x}dx  + \sum_{i\in\mathcal{I}_{\partial}} v_i  \int\displaylimits_{\Omega}\frac{\partial \psi_j}{\partial x}\frac{\partial \psi_i}{\partial x}dx \right) \\
&- \left(r_b- \frac{\sigma^2}{2}\right) \left( \sum_{i=1}^n v_i \int\displaylimits_{\Omega}\frac{\partial \psi_j}{\partial x}\psi_idx   + \sum_{i\in\mathcal{I}_{\partial}}^n v_i \int\displaylimits_{\Omega}\frac{\partial \psi_j}{\partial x}\psi_i dx\right) \\
&-r_b \left( \sum_{i=1}^n v_i \int\displaylimits_{\Omega}\psi_j\psi_i dx + \sum_{i\in \mathcal{I}_{\partial}} v_i \int\displaylimits_{\Omega}\psi_j\psi_idx \right) + \mathcal{P}_{long,j}.
\end{aligned}
\label{eq: system of FEM for long}
\end{equation}
The global systems of equations~\eqref{eq: system of FEM for long} is assembled via local element matrices, whose structures depend on the choice of the basis functions $\psi_i$. The common choice for the basis functions is a class of functions satisfying the nodal condition,
\begin{eqnarray}
    \psi_i(x_j) = \begin{cases}
                         1,& i = j, \\
                         0,&\text{otherwise},
                    \end{cases} \label{eq:nodeq}
\end{eqnarray}
where $x_j$ is the nodal point. This choice leads to global systems of linear equations with tridiagonal and pentadiagonal coefficient matrices for P1 and P2 basis functions\footnote{ $n$ number of nodes in P2-FEM differs from P1-FEM being doubled. Like $n_{P2} = 2*\text{nE} + 1$, while $n_{P1} = \text{nE} + 1$. }, respectively. Unlike from~\cite{Forsyth2007,christara2022penalty}, the resulting matrices from nonlinear terms are not only tridiagonal but pentadiagonal whose inversion process will be time-consuming. However, the tradeoff between higher-order FEM and lower-order FDM is maintained in accuracy and the use of fewer degrees of freedom to get the desired solution, see the Section~\ref{sec: Numerical results}. Since the transformed PDE~\eqref{eq:HJB_transformed long} is defined in $x$-space, the solution is presented uniformly in $S$-space. To ensure consistency with the FDM benchmark results, which are based on a uniform mesh, we employed a non-uniform mesh with unequal element sizes.

\subsubsection{Linear Lagrangian basis function}
 Consider partition of the spatial domain $\Omega$ into $\text{nE}$ non-overlapping elements $\Omega_j = [x_{j-1},x_j]$, with $|\Omega_j| = x_j - x_{j-1} = h$, $x_j$, $j = 0,\dots, \text{nE}$, the nodal points, $x_0 = x_{\min}$ and $x_{\text{nE}} = x_{\max}$. In the basic element $\Omega_j = [x_{j-1},x_j]$, we define two linear interpolation basis functions on parametric coordinates, $\xi$, which will later be transformed to physical coordinates $x$
\begin{align}
  \phi_1(\xi)=\xi \quad \text { and } \quad \phi_2(\xi)=1-\xi, \quad 0 \leq \xi \leq 1,\notag
\end{align}
which are called the local linear shape functions in the parametric coordinate $\xi$ associated with $\phi_1^{(e)}(x)$ and $\phi_2^{(e)}(x)$. Where $$\xi(x)=\frac{x-x_1^{(e)}}{x_2^{(e)}-x_1^{(e)}}, \quad x \in\left[x_1^{(e)}, x_2^{(e)}\right].$$ The inverse of $\xi(x)$ is $x(\xi)=x_1^{(e)}+\left(x_2^{(e)}-x_1^{(e)}\right) \xi$.
After transformation, linear basis functions are: 
\begin{align}
    \psi_{j-1}(x) &= (x - x_j)/(x_{j-1} - x_j) = - (x-x_j)/h, \notag \\
    \psi_j (x)      &= (x-x_{j-1})/(x_j - x_{j-1}) = (x-x_{j-1})/h, \notag
\end{align}
Evaluating the integrals in~\eqref{eq: system of FEM for long} using the above-stated basis functions over the element $\Omega_j$ results in the following local (element) matrices:
\begin{itemize}
\item For the $\displaystyle \int \psi_j \psi_i dx$ term, the element matrix reads
\begin{equation*}
    \boldsymbol{M}_j =  \begin{bmatrix}
            \displaystyle  \int \displaylimits_{\Omega_{j}} \psi_{j-1} \psi_{j-1}dx & \displaystyle \int \displaylimits_{\Omega_{j}} \psi_{j-1} \psi_{j}dx \\
 \displaystyle \int \displaylimits_{\Omega_{j}} \psi_{j} \psi_{j-1}dx &\displaystyle  \int \displaylimits_{\Omega_{j}} \psi_{j} \psi_{j}dx
 \end{bmatrix} = \frac{h}{6} \begin{bmatrix}
    2 & 1 \\
    1 & 2
\end{bmatrix}.
\end{equation*}

\item For the $-\displaystyle \int \psi_{j,x} \psi_{i,x} dx$ term, the element matrix reads
\begin{equation*}
    \boldsymbol{K}_j =  -\begin{bmatrix}
         \displaystyle  \int \displaylimits_{\Omega_{j}} \psi_{j-1,x} \psi_{j-1,x}dx & \displaystyle \int \displaylimits_{\Omega_{j}} \psi_{j-1,x} \psi_{j,x}dx \\
 \displaystyle \int \displaylimits_{\Omega_{j}} \psi_{j,x} \psi_{j-1,x}dx &\displaystyle  \int \displaylimits_{\Omega_{j}} \psi_{j,x} \psi_{j,x}dx
  \end{bmatrix} = 
  -\frac{1}{h} \begin{bmatrix}
   1 & -1 \\
 -1 & 1
  \end{bmatrix}.
\end{equation*}

\item For the $\displaystyle \int \psi_{j} \psi_{i,x} dx$ term, the element matrix reads
\begin{equation*}
    \boldsymbol{N}_j =  \begin{bmatrix}
   \displaystyle  \int \displaylimits_{\Omega_{j}} \psi_{j-1} \psi_{j-1,x}dx & \displaystyle \int \displaylimits_{\Omega_{j}} \psi_{j-1} \psi_{j,x}dx \\
 \displaystyle \int \displaylimits_{\Omega_{j}} \psi_{j} \psi_{j-1,x}dx &\displaystyle  \int \displaylimits_{\Omega_{j}} \psi_{j} \psi_{j,x}dx
 \end{bmatrix} = 
 \frac{1}{2} \begin{bmatrix}
     -1 & 1 \\
      -1 & 1
 \end{bmatrix}.
\end{equation*}
\end{itemize}

\subsubsection{Quadratic Lagrangian basis function}
In this approach, we add a midpoint $x_{j-\frac{1}{2}} = (x_{j-1} + x_j)/2$ in the basic element $\Omega_j$, giving three nodal points: $x_{j-1}$, $x_{j-1/2}$, and $x_j$, and define three quadratic interpolation polynomials satisfying the nodal condition~\eqref{eq:nodeq} on parametric coordinates, $\xi$ which will later be transformed to physical coordinates $x$:
\begin{align}
   \phi_1(\xi)=\frac{-\xi(1-\xi)}{2}, \quad \phi_2(\xi)=(1-\xi)(1+\xi), \quad \phi_3(\xi)=\frac{\xi(1+\xi)}{2}\notag
\end{align}
resulting in P2-FEM. $$
x=\frac{x_3^{(e)}+x_1^{(e)}}{2}+\frac{\left(x_3^{(e)}-x_1^{(e)}\right)}{2} \xi
$$
be the linear transformation, and for simplicity also let $\displaystyle x_2^{(e)}=\frac{x_3^{(e)}+x_1^{(e)}}{2}$. Once the transformation is done, we have: 
\begin{align}
    \psi_{j-1}(x) &= \frac{(x-x_{j-\frac{1}{2}})(x-x_j)}{(x_{j-1} - x_{j-\frac{1}{2}})(x_{j-1} - x_{j})} =  2(x-x_{j-\frac{1}{2}})(x-x_j)/h^2, \notag\\
   \psi_{j-\frac{1}{2}}(x) &= \frac{(x-x_{j-1})(x-x_j)}{(x_{j-\frac{1}{2}} - x_{j-1})(x_{j-\frac{1}{2}} - x_{j})} = -4(x-x_{j-1})(x-x_j)/h^2, \notag\\
    \psi_{j}(x) &= \frac{(x-x_{j-1})(x-x_{j-\frac{1}{2}})}{(x_j - x_{j-1})(x_j - x_{j-\frac{1}{2}})} =  2(x-x_{j-1})(x-x_{j-\frac{1}{2}})/h^2,\notag
\end{align}
 The local element matrices are as follows:
\begin{itemize}
\item the $\displaystyle \int \psi_j \psi_i dx$ term:
\begin{equation*}
    \boldsymbol{M}_j =  \begin{bmatrix}
     \displaystyle  \int \displaylimits_{\Omega_{j}} \psi_{j-1} \psi_{j-1}dx & \displaystyle \int \displaylimits_{\Omega_{j}} \psi_{j-1} \psi_{j-\frac{1}{2}}dx& \displaystyle \int \displaylimits_{\Omega_{j}} \psi_{j-1} \psi_{j}dx \\
 \displaystyle  \int \displaylimits_{\Omega_{j}} \psi_{j-\frac{1}{2}} \psi_{j-1}dx & \displaystyle \int \displaylimits_{\Omega_{j}} \psi_{j-\frac{1}{2}} \psi_{j-\frac{1}{2}}dx& \displaystyle \int \displaylimits_{\Omega_{j}} \psi_{j-\frac{1}{2}} \psi_{j}dx \\
 \displaystyle \int \displaylimits_{\Omega_{j}} \psi_{j} \psi_{j-1}dx  &\displaystyle  \int \displaylimits_{\Omega_{j}} \psi_{j} \psi_{j-\frac{1}{2}}dx &\displaystyle  \int \displaylimits_{\Omega_{j}} \psi_{j} \psi_{j}dx
\end{bmatrix} = 
\frac{h}{30} \begin{bmatrix}
   4 & 2 & -1 \\
   2 & 16 & 2 \\
  -1 & 2  & 4
\end{bmatrix}.
\end{equation*}
\item the $\displaystyle -\int \psi_{j,x} \psi_{i,x} dx$ term:
\begin{equation*}
    \boldsymbol{K}_j = - \begin{bmatrix}
     \displaystyle  \int \displaylimits_{\Omega_{j}} \psi_{j-1,x} \psi_{j-1,x}dx & \displaystyle \int \displaylimits_{\Omega_{j}} \psi_{j-1,x} \psi_{j-\frac{1}{2},x}dx& \displaystyle \int \displaylimits_{\Omega_{j}} \psi_{j-1,x} \psi_{j,x}dx \\
 \displaystyle  \int \displaylimits_{\Omega_{j}} \psi_{j-\frac{1}{2},x} \psi_{j-1,x}dx & \displaystyle \int \displaylimits_{\Omega_{j}} \psi_{j-\frac{1}{2},x} \psi_{j-\frac{1}{2},x}dx& \displaystyle \int \displaylimits_{\Omega_{j}} \psi_{j-\frac{1}{2},x} \psi_{j,x}dx \\
 \displaystyle \int \displaylimits_{\Omega_{j}} \psi_{j,x} \psi_{j-1,x}dx  &\displaystyle  \int \displaylimits_{\Omega_{j}} \psi_{j,x} \psi_{j-\frac{1}{2},x}dx &\displaystyle  \int \displaylimits_{\Omega_{j}} \psi_{j,x} \psi_{j,x}dx
 \end{bmatrix} = 
 -\frac{1}{3h} \begin{bmatrix}
  7 & -8 & 1 \\
 -8 & 16 & -8 \\
 1 & -8  & 7
\end{bmatrix}.
\end{equation*}

\item the $\displaystyle \int \psi_{j} \psi_{i,x} dx$ term:
\begin{equation*}
   \boldsymbol{N}_j = \begin{bmatrix}
        \displaystyle  \int \displaylimits_{\Omega_{j}} \psi_{j-1} \psi_{j-1,x}dx & \displaystyle \int \displaylimits_{\Omega_{j}} \psi_{j-1} \psi_{j-\frac{1}{2},x}dx& \displaystyle \int \displaylimits_{\Omega_{j}} \psi_{j-1} \psi_{j,x}dx \\
 \displaystyle  \int \displaylimits_{\Omega_{j}} \psi_{j-\frac{1}{2}} \psi_{j-1,x}dx & \displaystyle \int \displaylimits_{\Omega_{j}} \psi_{j-\frac{1}{2}} \psi_{j-\frac{1}{2},x}dx& \displaystyle \int \displaylimits_{\Omega_{j}} \psi_{j-\frac{1}{2}} \psi_{j,x}dx \\
 \displaystyle \int \displaylimits_{\Omega_{j}} \psi_{j} \psi_{j-1,x}dx  &\displaystyle  \int \displaylimits_{\Omega_{j}} \psi_{j} \psi_{j-\frac{1}{2},x}dx &\displaystyle  \int \displaylimits_{\Omega_{j}} \psi_{j} \psi_{j,x}dx
 \end{bmatrix} = \frac{1}{6} \begin{bmatrix}
  -3 & -4 & 1 \\
  4 & 0 & -4 \\
 -1 & 4  & 3
\end{bmatrix}.
\end{equation*}
\end{itemize}

\subsubsection{The treatment of the nonlinear term}
The global system of fintie element equations consist of one nonlinear term $\mathcal{P}_{long,j}$ in~\eqref{eq: system of FEM for long}. One shall construct the specific treatment for this type of non-smooth functions' integration as they differ from other linear standard terms in~\eqref{eq: system of FEM for long}. One possible technique is to apply \textit{group} FE~\cite{fletcher1983group,kazbek2024pricing}. However, we design the discrete versions of those functions~\eqref{eq: alpha functions for short and long}  directly by FE. Suppose that the term $\displaystyle\alpha_{long} := \min \left\{\mathsf{L}_1 V, \mathsf{L}_2 V, 0\right\}$ is approximated by
$$
  \alpha_{long} = \sum_{i=1}^n \alpha_i \psi_i + \sum_{\mathcal{I}_{\partial}} \alpha_i \psi_i,
$$
each operator in $\alpha_{long}$~\eqref{eq: alpha functions for short and long} is discretized by Galerkin method, hence, for $w = \psi_j$, $j = 1,\dots,n$, it reads
$$\mathsf{L}_1 v_i = (r_l-r_b)\left( \sum_{i=1}^n v_i \int\displaylimits_{\Omega}\frac{\partial \psi_j}{\partial x}\psi_idx   + \sum_{i\in\mathcal{I}_{\partial}}^n v_i \int\displaylimits_{\Omega}\frac{\partial \psi_j}{\partial x}\psi_i dx - \sum_{i=1}^n v_i \int\displaylimits_{\Omega} \psi_j\psi_idx   - \sum_{i\in\mathcal{I}_{\partial}}^n v_i \int\displaylimits_{\Omega}\psi_j\psi_i dx\right),$$
$$\mathsf{L}_2 v_i = -(r_b-r_l + r_f)\left( \sum_{i=1}^n v_i \int\displaylimits_{\Omega}\frac{\partial \psi_j}{\partial x}\psi_idx   + \sum_{i\in\mathcal{I}_{\partial}}^n v_i \int\displaylimits_{\Omega}\frac{\partial \psi_j}{\partial x}\psi_i dx \right),$$
therefore,
\begin{align}
\displaystyle \mathcal{P}_{long,j} &= \int\displaylimits_{\Omega} \sum_{i=1}^n \alpha_i \psi_i + \sum_{\mathcal{I}_{\partial}} \alpha_i \psi_i dx = \begin{cases}0 & \text { if }\mathsf{L}_1 v_i\geq 0 \text { and }\mathsf{L}_2 v_i\geq 0, \\
\mathsf{L}_1 v_i  & \text { if }\mathsf{L}_1 v_i <0 \text { and }\mathsf{L}_1 v_i<\mathsf{L}_2 v_i, \\ 
\mathsf{L}_2 v_i & \text { if }\mathsf{L}_2 v_i<0 \text { and }\mathsf{L}_1 v_i \geq\mathsf{L}_2 v_i .\end{cases}
\end{align}
Each integral in the above-equation is evaluated element-wise, resulting in the local element tridiagonal or pentadiagonal matrix $\boldsymbol{P}_{long}$.

\subsection{Time Integration Scheme}\label{sec: Time integration}

So far, we have covered the discretization in spatial direction based on P1-FEM and P2-FEM. The system of global equations~\eqref{eq: system of FEM for long} shall be supplemented by some time integration method. The Crank-Nicolson method (when $\theta = 1/2$) is applied together with Rannacher-start adjustment~\cite{Rannacher1984}. Due to the non-smooth initial data, the standard time integration schemes might face with some instability or spurious oscillations~\cite{christara2022penalty,forsyth2002quadratic}. Firstly, we will rewrite the global system of finite element equations as differential algebraic equations (DAEs) in matrix form
\begin{align}
    \frac{\partial}{\partial \tau}(\boldsymbol{M} \bsv  + \hat{\bsb}_{M,v})
    &= -\frac{\sigma^2}{2}  \boldsymbol{K} \bsv  - \left( r_b - \frac{\sigma^2}{2} \right) \boldsymbol{N}  \bsv - r_b \boldsymbol{M}  \bsv - \boldsymbol{\beta}_1 (\bsv) , \label{eq:linear_BS_matrix_form}\\
    &+  \boldsymbol{P}_{long} := F_1(\bsv)\notag \\
\end{align}
where
\begin{align}
    \displaystyle \boldsymbol{\beta_1}(\bsv) &= \frac{\sigma^2}{2}  \bsb_{K,v} + \left(r_b - \frac{\sigma^2}{2}  \right)\boldsymbol{b}_{N,v}+r_b \bsb_{M,v} -  \bsb_{long},
\end{align}
is the boundary condition vector. Applying the $\theta$-scheme on~\eqref{eq:linear_BS_matrix_form} results in the system with $\Delta \tau = T/n_{\tau}$, and $n_{\tau}$ the number of time steps,
\begin{align}
  \boldsymbol{M}  \bsv^{m+1} + \hat{\bsb}_{M,v}^{m+1} - \boldsymbol{M}  \bsv^{m} -\hat{\bsb}_{M,v}^{m} = \theta\Delta\tau F_1(\bsv^{m+1}) + (1-\theta)\Delta\tau F_1(\bsv^{m}), \notag
\end{align}
or
\begin{align}
    A_{11} \bsv^{m+1} &= \widetilde{A}_{11}\bsv^m + \theta \Delta \tau \boldsymbol{\beta}_1^{m+1} +  (1-\theta) \Delta \tau \boldsymbol{\beta}_1^{m} + \hat{\bsb}^m_{M,v} - \hat{\bsb}^{m+1}_{M,v}, \label{theta_scheme: for HJB long}
\end{align}
where
\begin{align}
    A_{11} &= \boldsymbol{M}  + \theta \Delta \tau \left( \frac{\sigma^2}{2} \boldsymbol{K}  + \left(r_b - \frac{\sigma^2}{2}  \right)\boldsymbol{N}  + r_b \boldsymbol{M}  + \boldsymbol{P}_{long} \right) \notag \\
    \widetilde{A}_{11} &= \boldsymbol{M}  - (1-\theta) \Delta \tau \left( \frac{\sigma^2}{2} \boldsymbol{K}  + \left(r_b - \frac{\sigma^2}{2}  \right)\boldsymbol{N}  + r_b\boldsymbol{M}+ \boldsymbol{P}_{long}    \right). \notag
\end{align}
In the spirit of Newton-type algorithms, we solve the DAE~\eqref{theta_scheme: for HJB long} using the penalty-like algorithm which was applied first in~\cite{Forsyth2007} and then improved by~\cite{christara2022penalty}. As in any iterative methods, there has to be at least one stopping criteria which is usually tolerance ($tol$) of two consequent iterands. We go further using the second stopping criteria which ensures the small difference (by tolerance) between two consequent penalty-like matrices. It is noteworthy that this condition is found to be almost identical with the penalty matrix equivalence relation condition in~\cite{kazbek2024pricing}. If one decides to come up with speedy convergent solution over conventional Newton's or family of relaxation techniques, perhaps, she would prefer to follow the algorithm below    
\begin{algorithm}[H]

\begin{algorithmic}[1]
\Require input $\bsv^m$(pay-off);
\State set $\bsv^{m+1,0} \leftarrow \bsv^{m+1}$ and $\boldsymbol{P}_{long}^{m, 0}=\boldsymbol{P}_{long}\left(\bsv^m\right)$;
;
\For{$k = 1,2,\dots$ until convergence}
    \State compute $\bsv^{m+1,k}$ using~\eqref{theta_scheme: for HJB long};
    \If{$\displaystyle \max _j\left\{\frac{\left|v_j^{m, k}-v_j^{m, k-1}\right|}{\max \left(\text { scale },\left|v_j^{m, k}\right|\right)}\right\}< tol$} 
        \State\algorithmicbreak
    \EndIf
    \State compute $\boldsymbol{P}_{long}^{m, k} = \boldsymbol{P}_{long}\left(\bsv^{m,k}\right)$ 
    \If{$\displaystyle \max _j\left\{\frac{\left|\left[P^{m, k-1} v^{m, k}-P^{m, k} v^{m, k}\right]\right|}{\max \left(\text { scale },\left|\left[P^{m, k} v^{m, k}\right]_j\right|\right)}\right\}<tol $} 
        \State\algorithmicbreak
    \EndIf
\EndFor
\State set $\bsv^{m }\leftarrow \bsv^{m+1,k}$
\end{algorithmic}
\caption{Computing the interior solutions}\label{alg:Interior}
\end{algorithm}

\section{Numerical Results}\label{sec: Numerical results}
In this section, one can find the numerical results and discussion for SBF for European straddle options. Temporal and spatial discretizations are realized as described in Sections~\ref{sec: Time integration} and~\ref{sec: Finite element method}. The price of Long and Short position straddle options with SBF are numerically computed utilizing the Galerkin FEM with linear and quadratic Lagrangian basis functions. The numerical results matches with benchmark range found in literature as well as the Greek letter values. 
\subsection{Example 1: Long position}
\begin{figure}[H]
    \centering
    \includegraphics[width=0.49\textwidth]{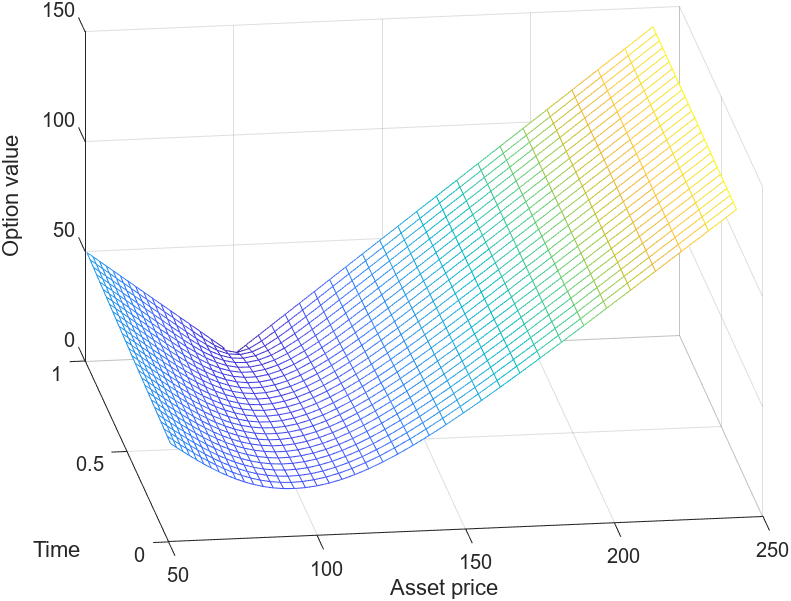}
    \includegraphics[width=0.49\textwidth]{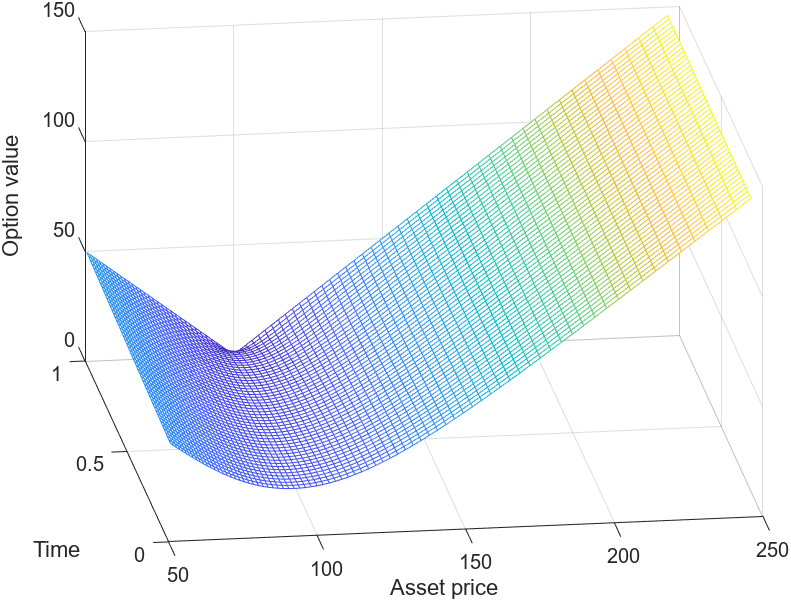}
    \caption{Surface solution of SBF for Long position by FDM with the following parameters: \(T = 1\), \(S_{\max} = 1000\), \(K = 100\), \(\sigma = 0.3\), \(r_b = 0.05\), \(r_l = 0.03\), \(r_f = 0.004\); Left figure: $\text{n}_{\text{FDM}}$ = 100, $N_t$ = 27; 
    Right figure: $\text{n}_{\text{FDM}}$ = 200, $N_t$ = 52;}
    \label{fig: 3D SBF Long FDM}
\end{figure}

\begin{figure}[H]
    \centering
    \includegraphics[width=0.49\textwidth]{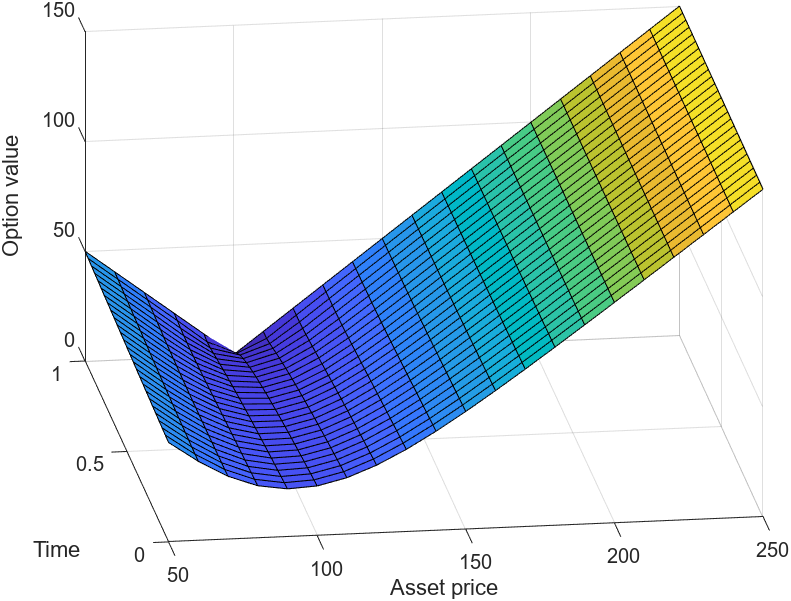}
    \includegraphics[width=0.49\textwidth]{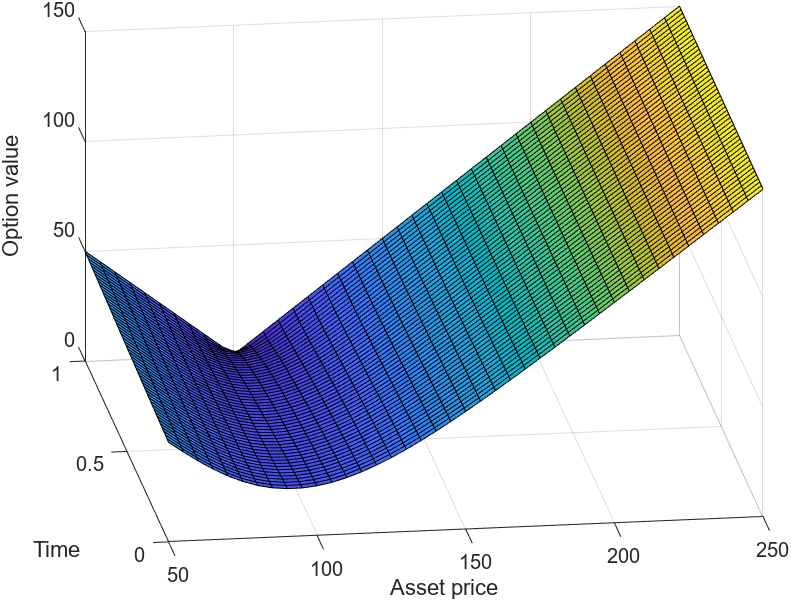}
    \caption{Surface solution of SBF for Long position by P1-FEM with the following parameters: \(T = 1\), \(S_{\max} = 1000\), \(K = 100\), \(\sigma = 0.3\), \(r_b = 0.05\), \(r_l = 0.03\), \(r_f = 0.004\); Left figure: nE = 100, $N_t$ = 27; 
    Right figure: nE = 200, $N_t$ = 52;}
    \label{fig: 3D SBF Long P1-FEM}
\end{figure}

\begin{figure}[H]
    \centering
    \includegraphics[width=0.49\textwidth]{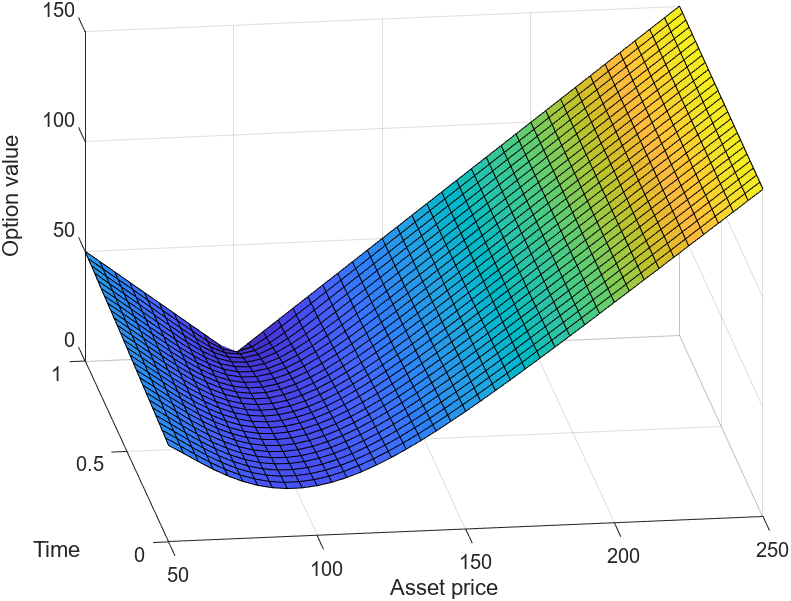}
    \includegraphics[width=0.49\textwidth]{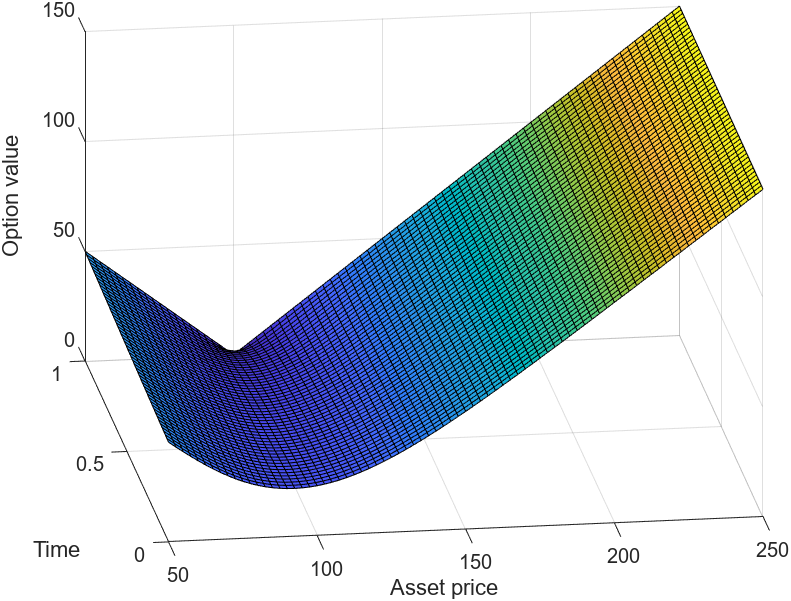}
    \caption{Surface solution of SBF for Long position by P2-FEM with the following parameters: \(T = 1\), \(S_{\max} = 1000\), \(K = 100\), \(\sigma = 0.3\), \(r_b = 0.05\), \(r_l = 0.03\), \(r_f = 0.004\); Left figure: nE = 100, $N_t$ = 27; 
    Right figure: nE = 200, $N_t$ = 52;}
    \label{fig: 3D SBF Long P2-FEM}
\end{figure}

\begin{figure}[H]
    \centering
    \includegraphics[width=0.49\textwidth]{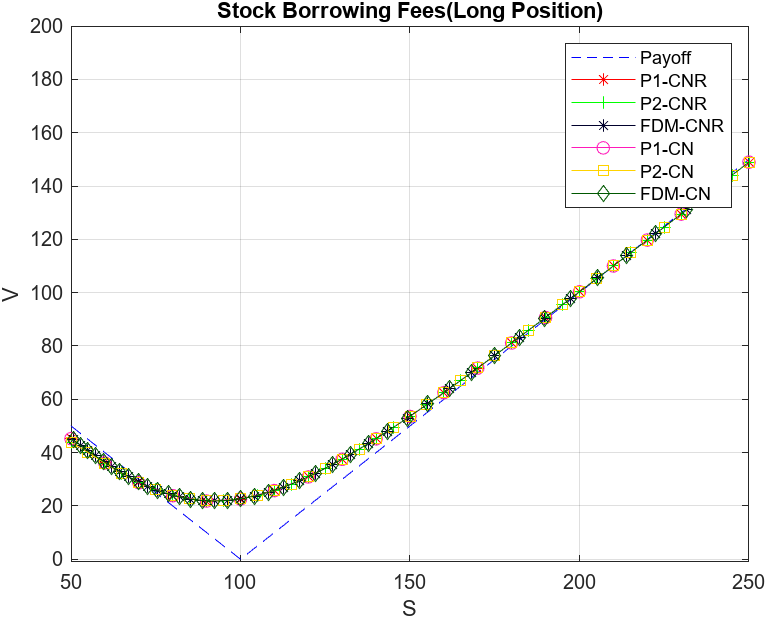}
    \includegraphics[width=0.49\textwidth]{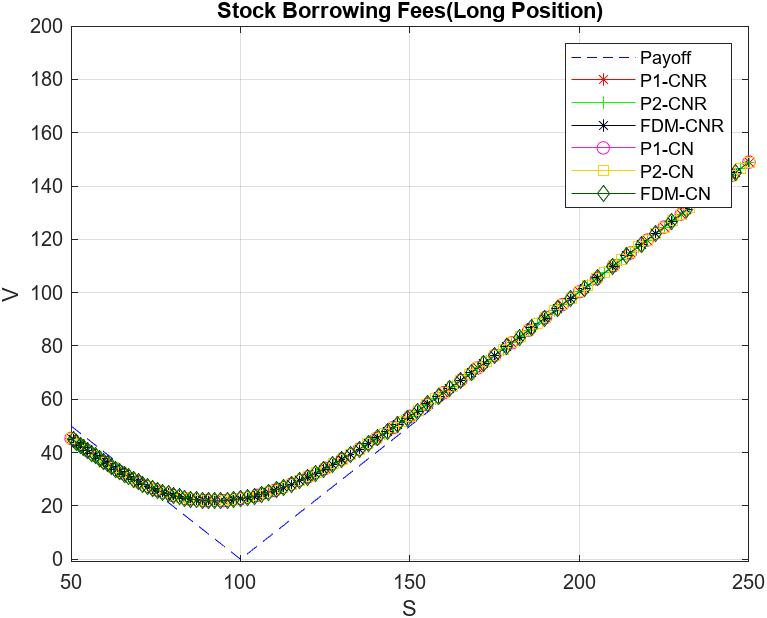}
    \caption{Solution of SBF for Long position at $t = 0$ with the following parameters: \(T = 1\), \(S_{\max} = 1000\), \(K = 100\), \(\sigma = 0.3\), \(r_b = 0.05\), \(r_l = 0.03\), \(r_f = 0.004\);
    Left figure: nE = $\text{n}_{\text{FDM}}$=100, Nt = 27; 
    Right figure: nE = $\text{n}_{\text{FDM}}$=200, Nt = 52.}
    \label{fig: 2D SBF Long comparison}
\end{figure}

We converted the semi-discrete FEM system~\eqref{eq: system of FEM for long} into fully discrete system of differential algebraic equations (DAE)~\eqref{theta_scheme: for HJB long} using the $\theta$-scheme. Now, we are at the position to present the numerical results for Long and Short position SBF task for European straddle option. We follow the standard stability condition, the chosen number of discretization in time and space agrees with the standard CFL condition for linear convection diffusion equation~\cite{kazbek2024pricing}. However, the conventional theoretical results for linear problems could be irrelevant for nonlinear problems. Numerical experiments with several refinement levels are producing no unexpected dynamic on the solution as we see from Figures~\ref{fig: 3D SBF Long FDM},~\ref{fig: 3D SBF Long P1-FEM}, and~\ref{fig: 3D SBF Long P2-FEM}. For the sake of fair comparison with the benchmark results, we reproduced them in our workstation. Otherwise, the relative execution timing and CPU time of different methods will be confusing as the timing of specific tasks are mattering to discuss because of many details such as the way of writing the program, the workstation capabilities, the difference between languages etc. In Figure~\ref{fig: 2D SBF Long comparison}, one can observe the comparison between FDM and FEM. A pure $\theta =1/2$-scheme is also used to compare with the Rannacher start version of it. 
In variable timestepping case, CNR is implemented to reduce the possible spurious oscillation that might come from the pay-off function~\cite{christara2018analysis}. As it was depicted along the smooth solution line, all methods are in good visual agreement among each other, with or without Rannacher adjustment. Henceforth we kept the numerical simulations with CNR-type time integration. In Table~\ref{tab: FDM for Long}, we reproduced the results from~\cite{Forsyth2007} to compare with FEM results. The adapted and revisited penalty-like iteration~\cite{christara2022penalty,Forsyth2007} is utilized to accelerate the computational time and reduce the number of iterations per each time-step needed to converge. Conventional penalty iteration shall involve the penalty parameter in American-style contracts~\cite{forsyth2002quadratic,deFrutos2005,Zvan1999}. As we deal with European-style exercise rights, we utilize the adapted version of penalty-iteration of Newton-type (Algorithm~\ref{alg:Interior}). FDM results paired with CNR are producing notable 4th order convergence ratio and matches with the reference results. Theoretically, Crank-Nicolson is of second order convergence, as well as central FD scheme. Additional superiority can be associated with the Rannacher time-stepping which smoothed the irregularity in initial data and led to higher order convergence. Similar performance is done with the P1-FEM results in Table~\ref{tab: P1-FEM for Long}. It is well known fact that for linear parabolic equation, the order of convergence according to a priori error estimates for linear test function is of second order, see e.g.,~\cite{oden2011introduction,kazbek2024phdthesis}. P1-FEM results are in good agreement with the reference results being consistent up to 4 decimal place accuracy. At this point, we did not observe any advantages of P1-FEM over classical FDM.

Another noteworthy observation related to the number of iterations per time step is that it is almost similar to the number of $N_t$, so the nonlinearity was treated efficiently by the Algorithm~\ref{alg:Interior} both in Tables~\ref{tab: P1-FEM for Long} and~\ref{tab: P2-FEM for Long}. In Table~\ref{tab: P2-FEM for Long}, we see that the results by P2-FEM and CNR are in good matching with the benchmark range and outperforms in terms of convergence properties by achieving the required accuracy at very early refinement level and maintaining the stability throughout. Note that, P2-FEM is computationally more expensive than P1-FEM or FDM, however, using notably lesser number of nodes it is offering  a more accurate solution.In Table~\ref{tab: P2-FEM for Long}, the ratio of P2-FEM we can see the dynamics in the convergence, which is quite unstable. Altough, it happens after sixth' decimal place as P2-FEM has converged at very early refinement levels than those by P1-FEM or FDM. The high-resolution numerical schemes, such as flux limiters, or SUPG (in FEM context) are not implemented for the current task, as we observed instabilities only after sevenths decimal place. Thereby, we maintain the monotonic convergence within five fractional digits for P2-FEM and fully monotone results for FDM and P1-FEM. 
\begin{table}[H]
    \centering
    \caption{Solution of SBF for Long position at $t = 0$ and $K = 100$ by FDM and CNR with the following parameters: \(T = 1\), \(S_{\max} = 1000\), \(K = 100\), \(\sigma = 0.3\), \(r_b = 0.05\), \(r_l = 0.03\), \(r_f = 0.004\);}
    \vspace{7pt}
    \begin{tabular}{c c c cc c c}
        \hline
        \multicolumn{5}{c}{Common Information} & \multicolumn{2}{c}{Iterations Information} \\
        \hline
        $\text{n}_{\text{FDM}}$ & $N_t$ & Value & Change & Ratio & Total & Average  \\ 
        \hline
        
         
         100 & 27 & 22.6358296529 & - & -&31 & 1.19 \\
         200 & 52 & 22.6722812919 & 0.0364516390 & -& 58 & 1.14 \\
         400 & 102 & 22.6813741266 & 0.0090928347 & 4.01 & 112 & 1.11 \\
         800 & 202 & 22.6836474618 & 0.0022733352 & 4.00 & 220 & 1.09 \\
         1600 & 402 & 22.6842160422 & 0.0005685804 & 4.00 & 436 & 1.09  \\
         3200 & 802 & 22.6843582547 & 0.0001422125 & 4.00 & 868 & 1.08 \\

    \end{tabular}
    \label{tab: FDM for Long}
\end{table}

    

\begin{table}[H]
    \centering
    \caption{Solution of SBF for Long position at $t = 0$ and $K = 100$ by P1-FEM and CNR with the following parameters: \(T = 1\), \(S_{\max} = 1000\), \(K = 100\), \(\sigma = 0.3\), \(r_b = 0.05\), \(r_l = 0.03\), \(r_f = 0.004\);}
    \vspace{7pt}
    \begin{tabular}{c c c c c c c c}
        \hline
        \multicolumn{5}{c}{Common Information} & \multicolumn{2}{c}{Iterations Information} & 
        \multicolumn{1}{c}{Relative Execution} \\
        \hline
        nE & $N_t$ & Value & Change & Ratio &  Total & Average & \(t_{P1-FEM}\)/\(t_{FDM}\) \\ 
        \hline
         100 & 27 & 22.5743839689 & - & - & 29 & 1.12 & 0.80 \\
         200 & 52 & 22.6581385915 & 0.0837546226 & - & 55 & 1.08 & 0.83 \\
         400 & 102 & 22.6779003316 & 0.0197617401 & 4.24 & 107 & 1.06 & 0.99 \\
         800 & 202 & 22.6827831398 & 0.0048828082 & 4.05 & 210 & 1.04 & 1.11\\
         1600 & 402 & 22.6839999628 & 0.0012168230 & 4.01 & 417 & 1.04 & 1.14\\
         3200 & 802 & 22.6843041310 & 0.0003041682 & 4.00 & 818 & 1.02 & 0.90 \\
    \end{tabular}
    \label{tab: P1-FEM for Long}
\end{table}

    

\begin{table}[H]
    \centering
    \caption{Solution of SBF for Long position at $t = 0$ and $K = 100$ by P2-FEM and CNR with the following parameters: \(T = 1\), \(S_{\max} = 1000\), \(K = 100\), \(\sigma = 0.3\), \(r_b = 0.05\), \(r_l = 0.03\), \(r_f = 0.004\);}
    \vspace{7pt}
    \begin{tabular}{c c c c c c c c}
        \hline
        \multicolumn{5}{c}{Common Information} & \multicolumn{2}{c}{Iterations Information} &
        \multicolumn{1}{c}{Relative Execution} \\
        \hline
        nE & $N_t$ & Value & Change & Ratio & Total & Average & \(t_{P2-FEM}\)/\(t_{FDM}\) \\ 
        \hline
         100 & 27 & 22.6841081248 & - & - & 35 & 1.35 & 0.61 \\
         200 & 52 & 22.6844496852 & 0.0003415604 & - & 61 & 1.20 & 0.85 \\
         400 & 102 &  22.6844047780 & 0.0000449072 & 7.60 & 116 & 1.15 & 0.78 \\
         800 & 202 &  22.6844044929 & 0.0000002851 & 157.51  &223 & 1.11 & 0.87 \\
         1600 & 402 & 22.6844053028 & 0.0000008099 & 2.84 & 431 & 1.07 & 0.86 \\
         3200 & 802 & 22.6844064552 & 0.0000011524 & 0.70  & 826 & 1.03 & 1.14
    \end{tabular}
    \label{tab: P2-FEM for Long}
\end{table}

\subsection{Example 2: Short position}
\begin{figure}[H]
    \centering
    \includegraphics[width=0.49\textwidth]{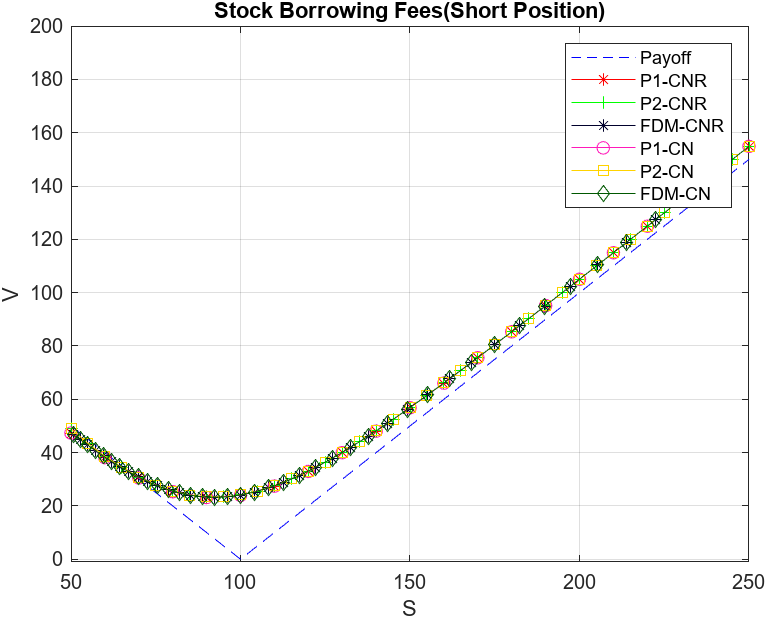}
    \includegraphics[width=0.49\textwidth]{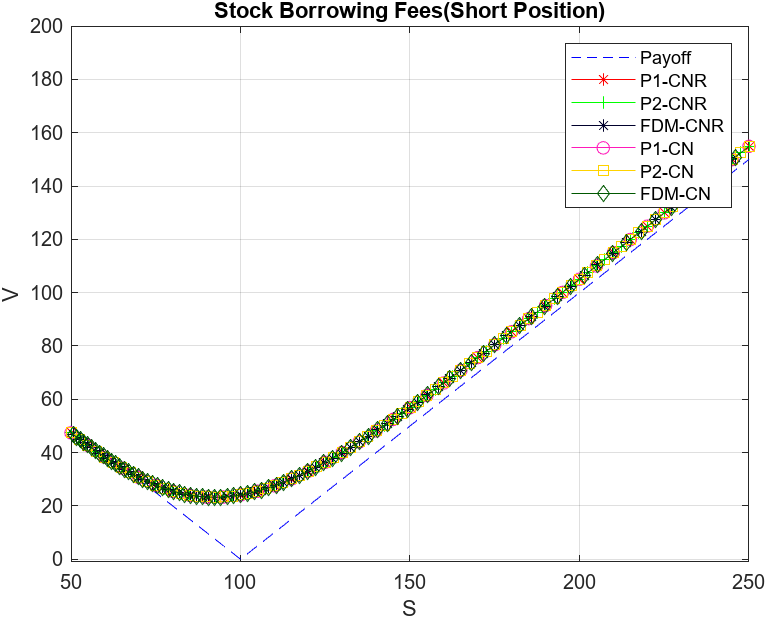}
    \caption{Solution of SBF for Short position at $t = 0$ with the following parameters: \(T = 1\), \(S_{\max} = 1000\), \(K = 100\), \(\sigma = 0.3\), \(r_b = 0.05\), \(r_l = 0.03\), \(r_f = 0.004\);
    Left figure: nE =$\text{n}_{\text{FDM}}$= 100, Nt = 27; 
    Right figure: nE = $\text{n}_{\text{FDM}}$=200, Nt = 52.}
    \label{fig: 2D SBF Short comparison}
\end{figure}

In Figure~\ref{fig: 2D SBF Short comparison}, the comparison is conducted among several different methodologies and any visual inconsistencies are not detected for early two refinement levels. The price of the short position is behaving almost in the same fashion as for the long position case. In Table~\ref{tab: FDM for Short}, the benchmark result~\cite{Forsyth2007} is reproduced too as before. P1-FEM and FDM agrees well with among each other being ultimately consistent in terms of relative-execution timing and ratios, in Tables~\ref{tab: FDM for Short} and~\ref{tab: P1-FEM for Short}. 

\begin{table}[H]
    \centering
    \caption{Solution of SBF for Short position at $t = 0$ and $K = 100$ by FDM and CNR with the following parameters: \(T = 1\), \(S_{\max} = 1000\), \(K = 100\), \(\sigma = 0.3\), \(r_b = 0.05\), \(r_l = 0.03\), \(r_f = 0.004\);}
    \vspace{7pt}
    \begin{tabular}{c c c c c c c}
        \hline
        \multicolumn{5}{c}{Common Information} & \multicolumn{2}{c}{Iterations Information} \\
        \hline
        $\text{n}_{\text{FDM}}$ & $N_t$ & Value & Change & Ratio &  Total & Average \\ 
        \hline
         100 & 27 & 24.0813519462 &  - & - & 32 & 1.23 \\
         200 & 52 & 24.1215308991 & 0.0401789529 & -  & 58 & 1.14 \\
         400 & 102 & 24.1313062926 & 0.0097753935 &  4.11 & 112 & 1.11  \\
         800 & 202 & 24.1337283777 & 0.0024220851 & 4.04 & 221 & 1.10  \\
         1600 & 402 & 24.1343307575 & 0.0006023798 & 4.02 &  439 & 1.09  \\
         3200 & 802 & 24.1344824664 & 0.0001517089 & 3.96 &  894 & 1.12\\
    \end{tabular}
    \label{tab: FDM for Short}
\end{table}
\noindent
The notable advantage of the former is that the discretized DAEs are performing better in terms of number of iterations spent for convergence than the latter. Slightly smaller average number of iterations were spent for each time-step than in FDM. We have to mention that the element matrices from P1-FEM results in tridiagonal matrix structure, while in FDM for no derivatives terms it is just diagonal matrix. Therefore, at initial stages, FDM is faster than P1-FEM, however, if we focus on the convergent price, they both converge by almost same speed. 

\begin{table}[H]
    \centering
    \caption{Solution of SBF for Short osition at $t = 0$ and $K = 100$ by P1-FEM and CNR with the following parameters: \(T = 1\), \(S_{\max} = 1000\), \(K = 100\), \(\sigma = 0.3\), \(r_b = 0.05\), \(r_l = 0.03\), \(r_f = 0.004\);}
    \vspace{7pt}
    \begin{tabular}{c c c c c c c c}
         \hline
        \multicolumn{5}{c}{Common Information} & \multicolumn{2}{c}{Iterations Information} & 
        \multicolumn{1}{c}{Relative Execution} \\
        \hline
        nE & $N_t$ & Value & Change & Ratio &  Total & Average & \(t_{P1-FEM}\)/\(t_{FDM}\) \\ 
        \hline
         100 & 27 & 23.9600574257 & - & - & 29 & 1.12 & 0.72\\
         200 & 52 & 24.0909333242 & 0.1308758985 & - & 55 & 1.08 & 0.99 \\
         400 & 102 & 24.1238676442 & 0.0329343200 & 3.97  & 106 & 1.05  & 1.07 \\
         800 & 202 & 24.1319249234 & 0.0080572792 & 4.09 & 208 & 1.03 & 0.86  \\
         1600 & 402 & 24.1338704998 & 0.0019455764 & 4.14 & 410 & 1.02 & 1.01 \\
         3200 & 802 & 24.1343691930 & 0.0004986932 & 3.90 & 806 & 1.01 & 1.04 \\ 
         
    \end{tabular}
    \label{tab: P1-FEM for Short}
\end{table}

\begin{figure}[H]
    \centering
    \includegraphics[width=0.49\textwidth]{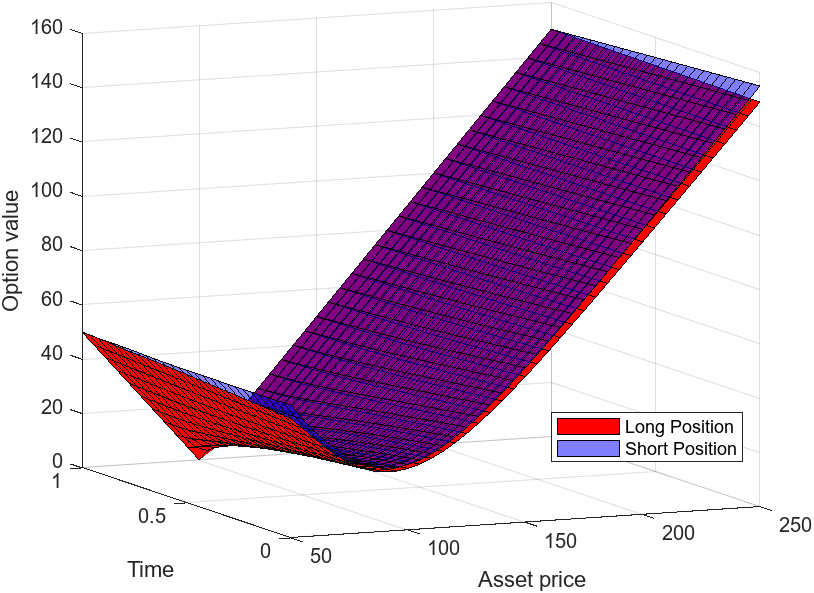}
    \includegraphics[width=0.49\textwidth]{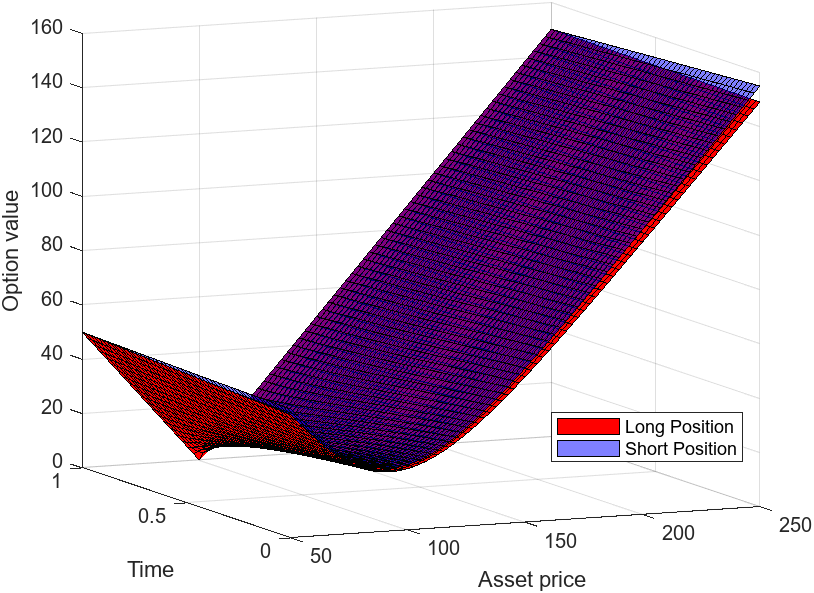}
    \caption{Surface solution comparison of SBF for Long and Short Positions by P2-FEM with the following parameters: \(T = 1\), \(S_{\max} = 1000\), \(K = 100\), \(\sigma = 0.3\), \(r_b = 0.05\), \(r_l = 0.03\), \(r_f = 0.004\);
    Left figure: nE = 100, Nt = 27; 
    Right figure: nE = 200, Nt = 52.}
    \label{fig: 3D SBF Long-Short comparison}
\end{figure}

\begin{table}[H]
    \centering
    \caption{Solution of SBF for Short position at $t = 0$ and $K = 100$ by P2-FEM and CNR with the following parameters: \(T = 1\), \(S_{\max} = 1000\), \(K = 100\), \(\sigma = 0.3\), \(r_b = 0.05\), \(r_l = 0.03\), \(r_f = 0.004\);}
    \vspace{7pt}
    \begin{tabular}{c c c c c c c c}
        \hline
        \multicolumn{5}{c}{Common Information} & \multicolumn{2}{c}{Iterations Information} & \multicolumn{1}{c}{Relative Execution} \\
        \hline
        nE & $N_t$ & Value & Change & Ratio &  Total & Average  & \(t_{P2-FEM}\) / \(t_{FDM}\) \\ 
        \hline
         100 & 27 & 24.1362186626 &  - & - & 33 &  1.27 & 0.91 \\
         200 & 52 & 24.1348073373 & 0.0014113253 & - & 61 & 1.20 & 0.92 \\
         400 & 102 & 24.1346048257 & 0.0002025116 & 6.97 & 115 & 1.14 & 0.66 \\
         800 & 202 & 24.1345181605 & 0.0000866652 & 2.34 & 219 & 1.09 & 0.78 \\
         1600 & 402 & 24.1345352543 & 0.0000170938 & 5.07 & 416 & 1.04 & 0.93 \\
         3200 & 802 & 24.1345333239 & 0.0000019304 & 8.86 & 813 & 1.01 & 1.20
    \end{tabular}
    \label{tab: P2-FEM for Short}
\end{table}
In Table~\ref{tab: P2-FEM for Short}, P2-FEM performs quite faster than those by P1-FEM or FDM, in terms of convergence. As it has spent only first two refinement levels to reach the same accuracy by other two methods. These results are quite consistent with the results presented for the Long position case, the fast convergence property has inherited from Long to Short position which validates the stability of the method for these specific two problems. Also, we show that the difference between Long and Short position cases are notable in terms of the price at time $t=0$ as can easily be seen from Figure~\ref{fig: 3D SBF Long-Short comparison}. The relative execution timing is computed by maintaining the fair comparison between nE and $N_t$. In fact, the time spent for the convergent result in FDM is $490$ seconds, whereas P2-FEM has spent $0.76$ seconds to get the same accuracy level for both Long and Short position cases. This result is convincing the usage of P2-FEM as an alternative method over other classical methods. Moreover, the monotonic convergence is secured within five fractional digits notably matching with the reference~\cite{Forsyth2007}.


\subsection{Greeks}
The reflection of the prices, its sensitivities, are described numerically using P2-FEM and FDM. It is well-known fact that the Greeks are of great important to facilitate the practical usage of these results for the portfolio rebalancing and hedging goals under no arbitrage assumptions. In this section, the Greek letters are compared between P2-FEM and FDM. The intention of this section is to suggest an alternative method for the Greek value computation. 
\begin{figure}[H]
    \centering
    \includegraphics[width=0.30\textwidth]{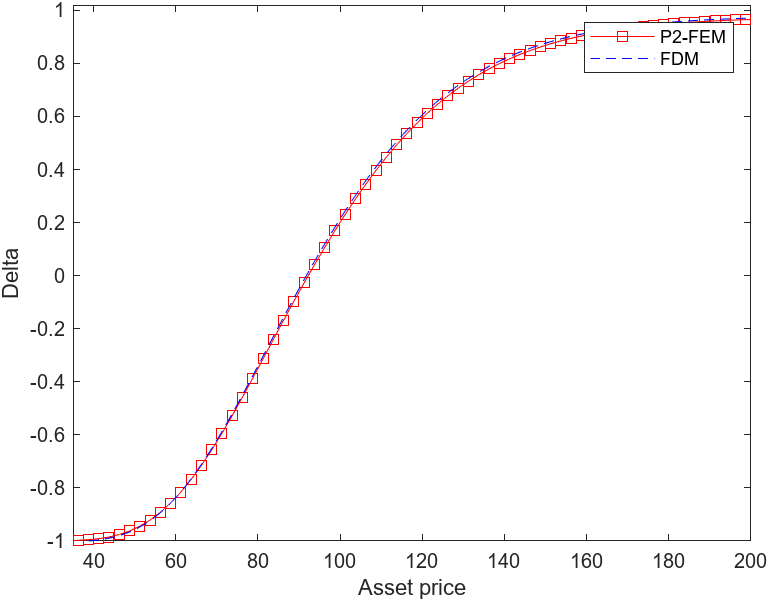}
    \includegraphics[width=0.30\textwidth]{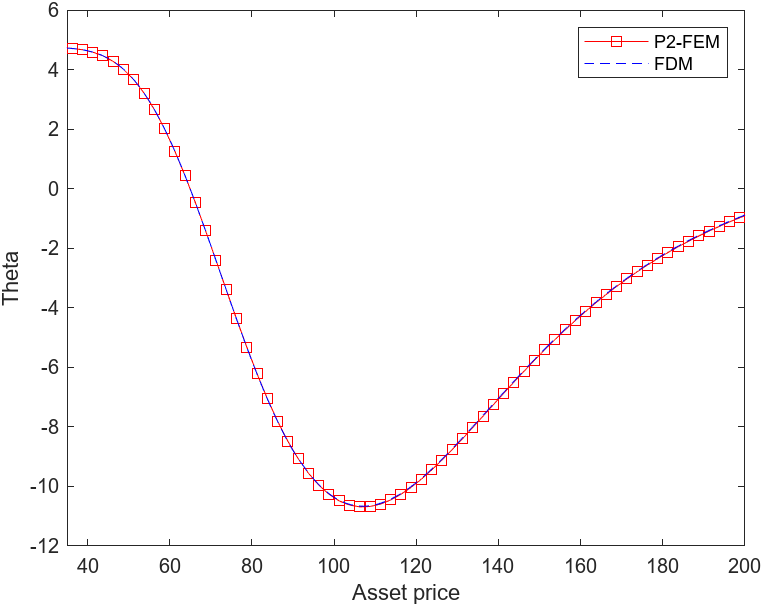}
    \includegraphics[width=0.30\textwidth]{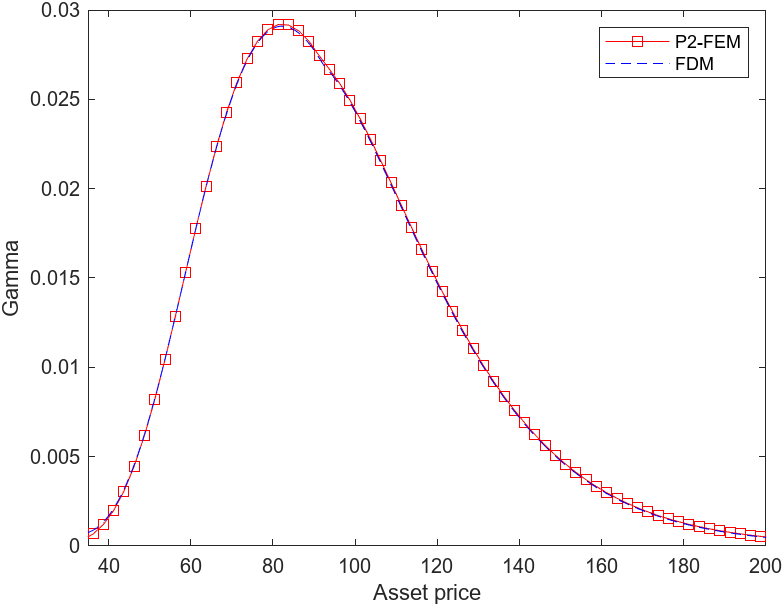}
    \caption{Greeks of SBF for Long position at \(t = 0\) with the following parameters: \(T = 1\), \(S_{\max} = 1000\), \(K = 100\), \(\sigma = 0.3\), \(r_b = 0.05\), \(r_l = 0.03\), \(r_f = 0.004\), nE = $\text{n}_{\text{FDM}}$=400, $N_t$ = 102;}
    \label{fig: 2D SBF Long  Greeks comparison}
\end{figure}

\begin{figure}[H]
    \centering
    \includegraphics[width=0.30\textwidth]{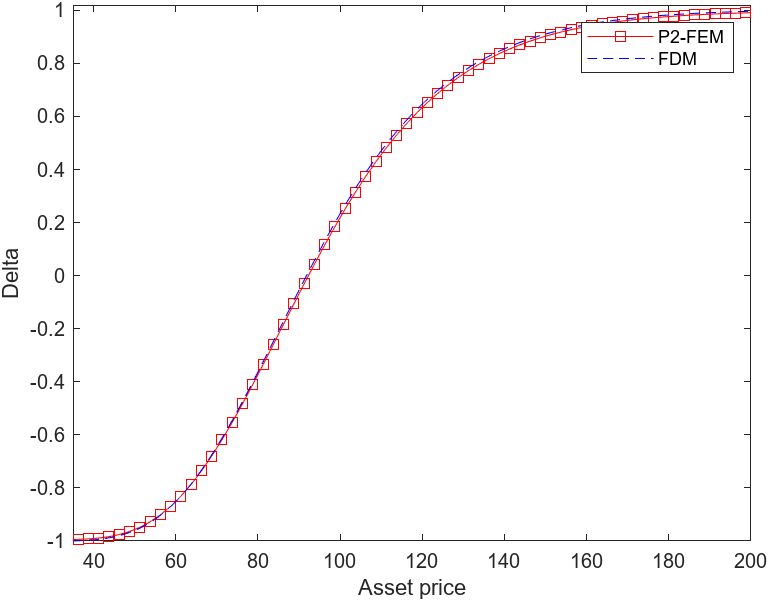}
    \includegraphics[width=0.30\textwidth]{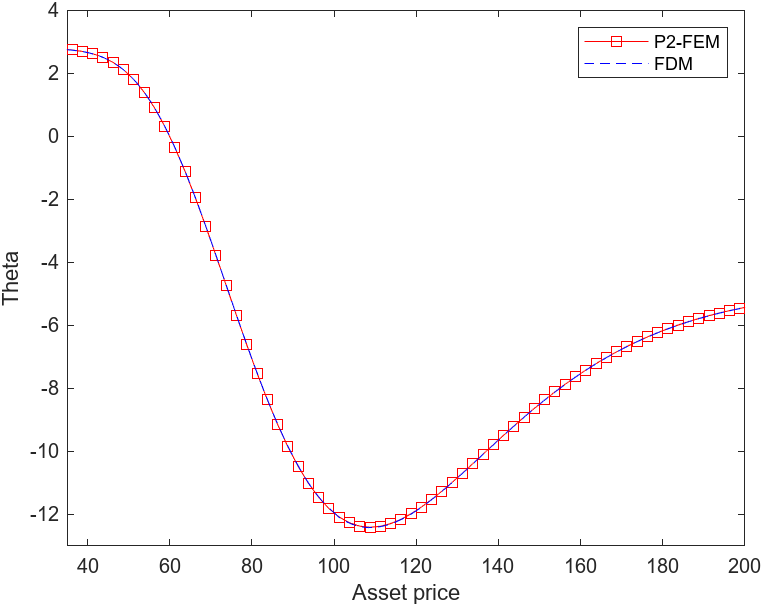}
    \includegraphics[width=0.30\textwidth]{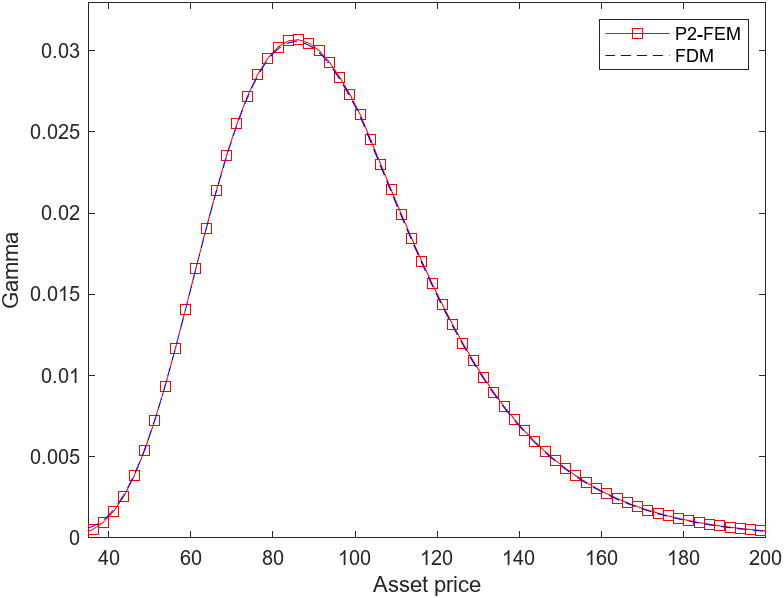}
    \caption{Greeks of SBF for Song position at \(t = 0\) with the following parameters: \(T = 1\), \(S_{\max} = 1000\), \(K = 100\), \(\sigma = 0.3\), \(r_b = 0.05\), \(r_l = 0.03\), \(r_f = 0.004\), nE=$\text{n}_{\text{FDM}}$ = 400, $N_t$ = 102;}
    \label{fig: 2D SBF Short  Greeks comparison}
\end{figure}
\noindent
We mainly refer the methodology part of the Greek computations to this work~\cite{kazbek2024pricing}, where we utilized the derived approximation formulas to obtain the corresponding Greek letters. Practically important thing about the Greeks is its smoothness, as for nonlinear PDE post-processing values it is a challenging task to obtain. The nonlinearity in HJB task for SBF appears in first-order derivatives being even more irritating, perhaps, as it affects to the convection term making the problem more unstable. In Figures~\ref{fig: 2D SBF Long  Greeks comparison},~\ref{fig: 2D SBF Short  Greeks comparison}, we see that the all possible Greek quantities for one-factor model is presented. As shown in Figure~\ref{fig: 3D SBF Long Greek Surfaces}, no spikes, shocks, or oscillations are observed in either the current time values or the surface values, and the Greek values compare favorably with those obtained using FDM.
\begin{figure}[H]
    \centering
    \includegraphics[width=0.49\textwidth]{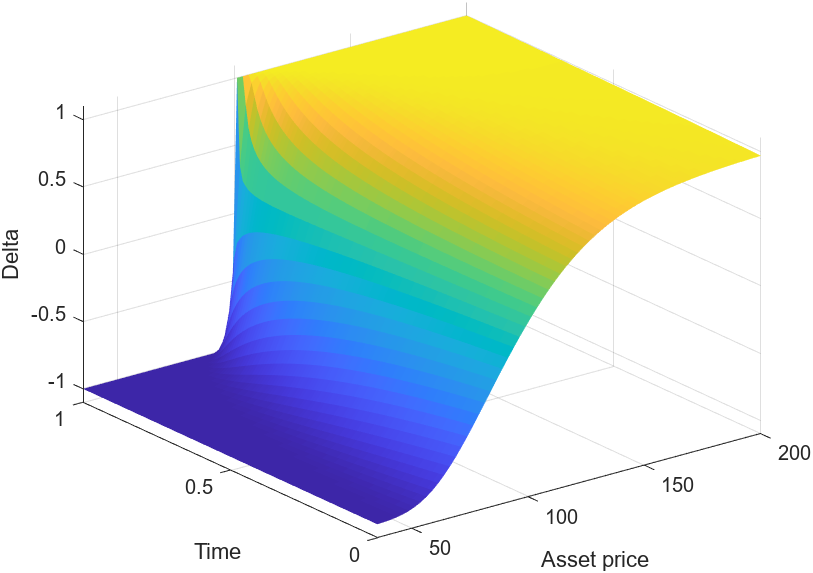}
    \includegraphics[width=0.49\textwidth]{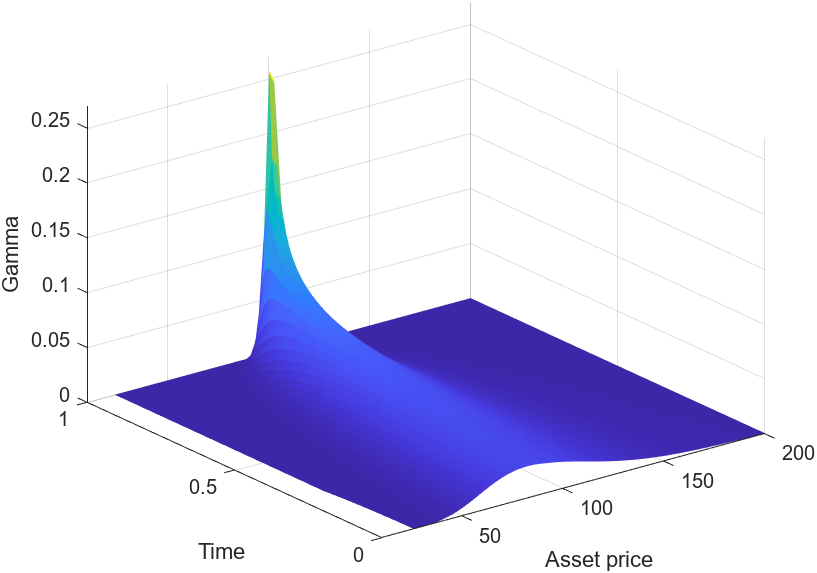}
    \includegraphics[width=0.49\textwidth]{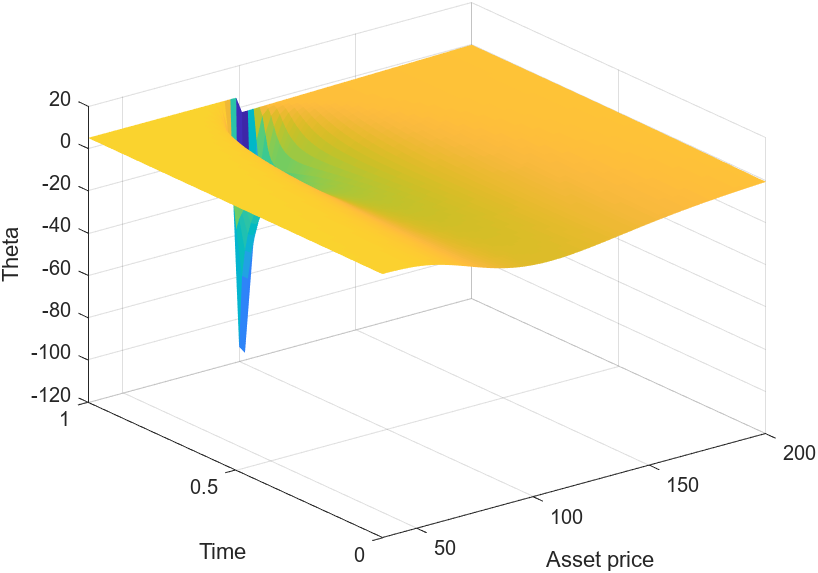}
    \caption{Greek surfaces of SBF for Long position by P2-FEM with the following parameters: \(T = 1\), \(S_{\max} = 1000\), \(K = 100\), \(\sigma = 0.3\), \(r_b = 0.05\), \(r_l = 0.03\), \(r_f = 0.004\), nE = 400, $N_t$ = 102;} 
    \label{fig: 3D SBF Long  Greek Surfaces}
\end{figure}

\section{Conclusion}
In this paper, we apply the Finite Element Method (FEM) on a nonuniform mesh to solve the HJB PDEs for pricing European straddle options. The element matrices are computed using the Galerkin method, based on the discretization of the subdomains. A semi-discrete system of equations is then fully discretized using the Crank-Nicolson-Rannacher time-stepping method. Nonlinear terms are accurately resolved at each time step through an iterative method, demonstrating fast convergence. Higher-order P2-FEM quadratic basis functions reveal notable performance improvements compared to benchmark results. The CPU time required to achieve the desired results shows a significant advantage for P2-FEM over FDM and linear P1-FEM, with P2-FEM demonstrating much faster convergence. We proposed an efficient alternative framework to the HJB problem and contributed to the literature by introducing an FEM-based solution for HJB tasks commonly encountered in mathematical finance. Although, the numerical examples presented are showcasing the existing benchmark properties, the solid theoretical a priori error estimate results supporting the experimental ones are still open question for the community. In part, the error estimates needs to be predicted for nonlinear HJB PDEs. Lastly, the Greek values are computed consistently via FEM and matched with FDM. Results on American-style Hamilton-Jacobi-Bellman-Isaacs PDEs will be presented in a forthcoming paper.
\section*{Declaration}
Everyone who contributed to this work was aware of the stage of the manuscript and agreed to submit it to the journal without any conflict of interests. All data and information used in this work are properly cited.
\section*{Data availability}
MATLAB P-codes for benchmarking and validation are available from the corresponding author upon reasonable request.
\section*{Funding}
This research was conducted without any financial support from external funding agencies, institutions, or organizations.

\bibliographystyle{unsrt}
\bibliography{main}
\end{document}